\documentclass[conference]{IEEEtran}
\IEEEoverridecommandlockouts
\usepackage{cite}
\usepackage{amsmath,amssymb,amsfonts}
\usepackage{algorithmic}
\usepackage{graphicx}
\usepackage{textcomp}
\usepackage{xcolor}
\def\BibTeX{{\rm B\kern-.05em{\sc i\kern-.025em b}\kern-.08em
    T\kern-.1667em\lower.7ex\hbox{E}\kern-.125emX}}
\begin{document}

\title{Blockchain-Enabled Secure Logging for Fiscal Electronic Mechanisms: Evaluation of the Greek “eSEND” and myDATA Tax Systems.\\
{\footnotesize \textsuperscript{}}
}

\author{\IEEEauthorblockN{Panagiotis Mavridis}
\IEEEauthorblockA{\textit{School of Engineering} \\
\textit{Dept. of Electronic Engineering}\\
Hellenic Mediterranean University \\
Chania, Greece \\
ORCID 0009-0006-7399-7659}
\and
\IEEEauthorblockN{Anargyros Baklezos, \IEEEmembership{Member, IEEE} }
\IEEEauthorblockA{\textit{School of Engineering} \\
\textit{Dept. of Electronic Engineering}\\
Hellenic Mediterranean University \\
Chania, Greece \\
ORCID 0000-0002-0532-6216}
\and
\IEEEauthorblockN{Christos Nikolopoulos, \IEEEmembership{Member, IEEE} }
\IEEEauthorblockA{\textit{School of Engineering} \\
\textit{Dept. of Electronic Engineering}\\
Hellenic Mediterranean University \\
Chania, Greece \\
ORCID 0000-0003-1344-4666}
}

\maketitle

\begingroup
\renewcommand{\thefootnote}{}
\footnotetext{%
  \raggedright\scriptsize
  \copyright~2026 IEEE. Personal use of this material is permitted.
  Permission from IEEE must be obtained for all other uses, in any current
  or future media, including reprinting/republishing this material for
  advertising or promotional purposes, creating new collective works,
  for resale or redistribution to servers or lists, or reuse of any
  copyrighted component of this work in other works.%
}
\addtocounter{footnote}{-1}
\endgroup
\begin{abstract}
This paper analyzes the implementation of blockchain-based integrity mechanisms in Greek Fiscal Electronic Mechanisms (FEMs) and the central tax information system “eSEND”. The study examines the cryptographic architecture of fiscal devices, including Electronic Cash Registers, Fiscal Printers, Fiscal Signing Machines, and FEMAS devices, which implement double or triple hash-chain structures to ensure transaction immutability. The transmission protocol between fiscal devices and the central database is also evaluated with respect to encryption, sequential validation, and blockchain verification. In contrast, the architecture of Electronic Invoicing Provider Services and the “myDATA” central platform is analyzed, highlighting the absence of blockchain-based integrity guarantees. The comparison demonstrates that hardware-based fiscal mechanisms provide stronger guarantees for transaction completeness and tamper resistance than purely software-based invoicing infrastructures. The findings highlight architectural weaknesses in the current e-invoicing framework and propose improvements for ensuring transaction integrity in digital tax ecosystems.
\end{abstract}


\section{Introduction}
A blockchain is a list of blocks that are securely connected together via hashes \cite{b1}. Each block contains a cryptographic hash of the previous block, a timestamp (date, time), and transaction data. Since each block contains information about the previous block, they form a chain, with each additional block linking to the previous block. Consequently, blockchain transactions are irreversible: once recorded and the cryptographic hash is computed, the data in any given block cannot be altered without altering all subsequent blocks.
Block chaining technology has been used for the authentication process of invoices, receipts, and other fiscal documents in various countries. The implementation of blockchaining for VAT statements of Danish companies is described in \cite{b5}.
In \cite{b3}, a block chaining for tax reporting is described, and its implementation is proposed for various fiscal documents. In \cite{b4}, block chaining implementation is described as a tool for Tax information exchange. 
In \cite{b5},  permissionless and permissioned blockchains for databases containing tax data are described. In the same article, it is demonstrated that although blockchains can facilitate the storage and sharing of tax information, they cannot streamline reporting requirements nor enhance cooperation between tax authorities. In \cite{b6}, it is demonstrated that implementing a blockchain-based platform for tax administration, would lead to digitalization and automation of various tax processes, would also increase the transparency of tax-related transactions and would reduce costs and other inefficiencies. In \cite{b7}, blockchain is studied in tax administration, as it can deliver reliable, real-time information from many sources to a large audience, resulting to an increase of efficiency and transparency. 
In \cite{b8}, the transition from paper-based tax system to a fully digital –based one, using permissioned blockchains is studied.
In \cite{b9}, the Intra Community Missing Trader “MTIC” fraud is presented as a fraud of omitting a VAT payment.  A good is imported, bought and consequently sold along a chain of companies. This good is finally exported. The first company who imported the good, collects VAT, but does not file a return off the tax collected and disappears. By using blockchain technology for the invoices issued, the “MTIC” fraud can be discovered prior the disappearance of the first company.
Several countries, including Brazil, Mexico, Hungary, Italy, China have forced companies to obtain digital stamps of approval of their invoices, so that to ensure VAT settlement. Digital stamps is a means of fulfilling the requirement of the authentication of the content and of the origin of any invoice at the time of issuing the invoice, which is a legal requirement imposed by European Directive 2006/112/EC  \cite{b10}  in article 246, and by the Greek Law 4308/2014 \cite{b11} in article 15.1. The most easy implementation of the above legal requirement is by obtaining digital stamps or hash values of the invoices and of the receipts, at the moment of their issuing. 
In this paper, authors describe a block-chaining implementation for securing transaction data in terms of hash-chain–based integrity schemes in Greek Fiscal Electronic Mechanisms “FEMs” e.g. Greek Fiscal Electronic Cash Registers ‘ECRs’, Fiscal Printers ‘AFDs’, Fiscal Signing Machines ‘EAFDSS’, Fiscal Electronic Mechanisms of Accumulation and Signing ‘FEMAS’ and EFTPOS. 
After securing transaction data using block chaining, these machines transmit this data to the central information system “eSEND”, where after confirming that the received data follows the block chaining rules, the server accepts it and stores it in its central database. The central information system “eSEND” operates under the responsibility of the Greek Independent Authority for Public Revenue “IAPR”.
We will also examine the operation of Electronic Invoicing Provider Services “EIPS” and the transmission of transaction data to the central database "myDATA", where we will note the absence of blockchain implementation and the consequences of this absence.
 
\section{Functionality of Greek FEMs}
\subsection{Fiscal Electronic Cash Registers: Triple Blockchain Implementation}

The technical specifications for the operation of Greek Fiscal Electronic Cash Registers “ECRs” are defined in POL 1135/2005 \cite{b12}, POL 1220/2012 \cite{b13}   and its amendments A.1024/2020 \cite{b14} and A.1173/2022 \cite{b15}. The obligations of FEM holders are specified in Law 4308/2014 and in POL 1068/2015. \cite{b16}.  The Fiscal ECRs operate within a package sealed by a fiscal screw, which allows access only to authorized technicians.
The working memory of Fiscal ECRs, consists of counters and accumulators, where receipt data are accumulated. They have also a built-in or removable SD memory card where all receipts are stored.

\begin{figure}[htbp]
\centering
\scalebox{.25}{\includegraphics{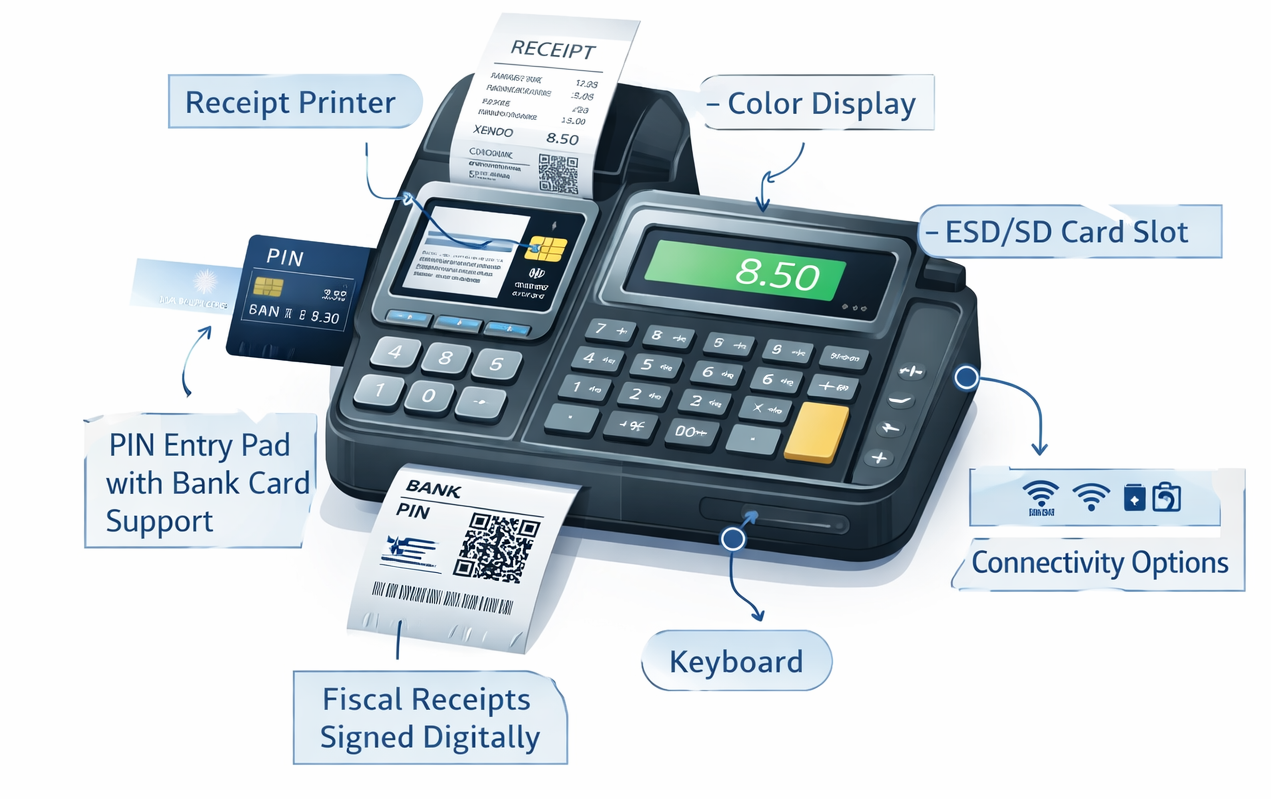}}
\caption{A greek restaurant-type Fiscal ECR.}
\label{fig1}
\end{figure}

\begin{figure}[htbp]
\centering
\scalebox{.1}{\includegraphics{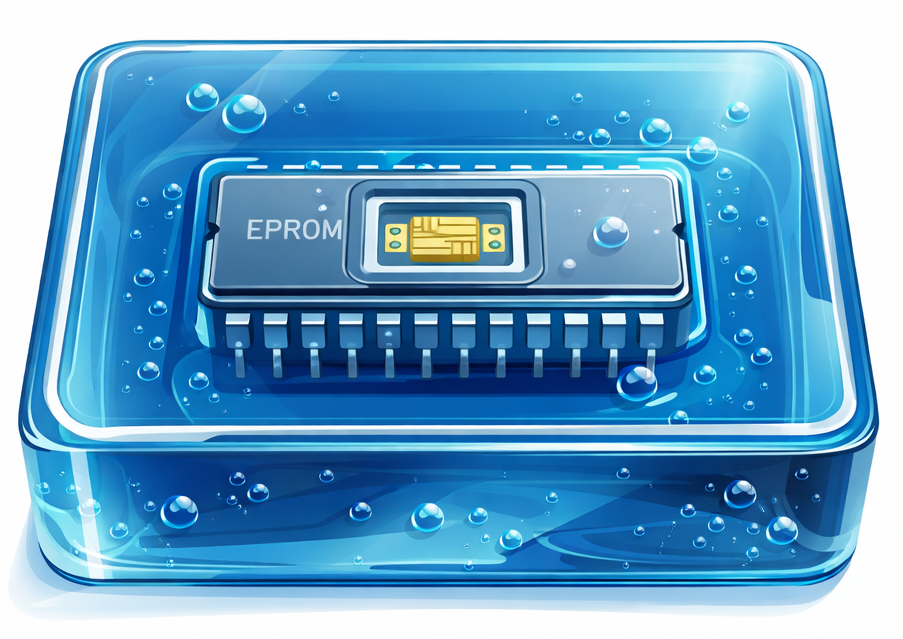}}
\scalebox{.2}{\includegraphics{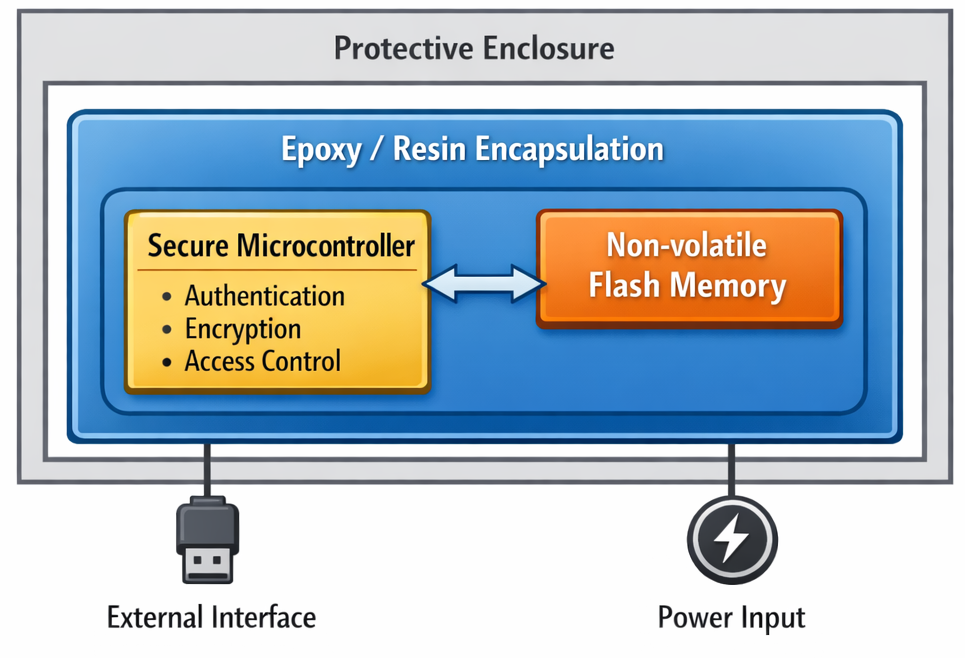}}
\caption{The fiscal memory, an EPROM in resin on the left, and the protection logic on the right.}
\label{fig2}
\end{figure}

The fiscal memory is typically an EPROM (Fig.~\ref{fig2}), or other type of memory controlled by a  microprocessor, where writing new data is allowed,  but deleting or modifying data is prohibited. The fiscal memory, for greater security, is located within a separate compartment and is surrounded by a special resin.  
In the working memory of the tax cash registers, the net value and VAT amounts per VAT category, resulting from each issued receipt, are added up.

In addition to the above memories depicted in Fig.~\ref{fig2}, each fiscal cash register has an SD card, built-in or removable depending on the model, where each receipt is recorded.  The original text of the receipt is placed in a file with the extension a.txt, which is recorded on the SD card.  When a receipt is issued, a hash value is calculated using the SHA-1 algorithm \cite{b17}, which is recorded in a corresponding file with the extension b.txt on the SD card. 
Upon issuing a new receipt, the fiscal ECR also creates a file with the extension e.txt which is also recorded on the SD card. The e.txt file consists of specific fields of fiscal interest defined in Table A of POL 1220/2012 .
At the end of the day, a “Z” report is issued. In fact, Greek Fiscal ECRs have an internal timer that blocks any functionality of the ECR 24 hours after issuing the first receipt, and requires issuing the “Z” report. 
Upon issuing the “Z” report, all amounts accumulated in the working memory are transferred from the accumulators to a new entry in the fiscal memory so that the working memory can accommodate the new data.  At the same time, corresponding progressive sums per VAT category are computed, as sums of the previous “Z” values, plus the sums of the present “Z” report.  In this way, a block chain is constructed, which is derived by the progressive amounts of each VAT category from each “Z” to the next “Z”.
Furthermore, when the Z report is issued, a new hash value is calculated using the SHA-1 algorithm that has as arguments the hash values of the b.txt files of the current "Z" receipts, as well as the hash value of the previous Z. This new hash value is stored in a new separate file on the SD card. The name of this file is derived from the serial number of the Fiscal Electronic Cash Register and the number of the “Z”, and ends in "c.txt".\\

\begin{equation}
c_{\text{txt}} =
\mathrm{SHA1}\!\left(
\sum_{i=1}^{n} H(b_{\text{txt}}^{(i)}, Z_{\text{txt}}^{\text{current}})
\;+\;
c_{\text{txt}}^{Z-1}
\right)
\label{eq:dtxt}
\end{equation}

Furthermore, when a “Z” report is issued, an “s.txt” file is constructed, that contains many lines, one line per receipt. Each line is constructed by multiple fields according to a specific order. The fields are semicolon delimited. The fields are: issuer VAT number, ECR s/n, various information, date and time, daily receipt number, ECR progressive receipt number, Z number, net value A, net value B, net value C, net value D, net value E, VAT amount A, VAT amount B, VAT amount C, VAT amount D, currency id, hash value.

\begin{figure}[htbp]
\centering
\includegraphics{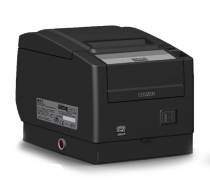}
\caption{The Fiscal Printer.}
\label{fig}
\end{figure}

Please note that the last field is a hash value derived from the contents of the previous fields. The last line of the “s.txt” file consists only from one field, which is the hash value of the hash values of the previous lines plus of the total hash value of the “s.txt” file of the previous “Z” report.  This hash value is referred as a “d.txt”.\\

\begin{equation}
d_{\text{txt}} =
\mathrm{SHA1}\!\left(
\sum_{i=1}^{n} H(e_{\text{txt}}^{(i)}, s_{\text{txt}}^{\text{current}})
\;+\;
d_{\text{txt}}^{Z-1}
\right)
\label{eq:dtxt}
\end{equation}

In this way, two more block chains of hash values are constructed, one with the “c.txt” hash values and another with the “d.txt” hash values.

\begin{figure*}[htbp]
\centering
\scalebox{.45}{\includegraphics{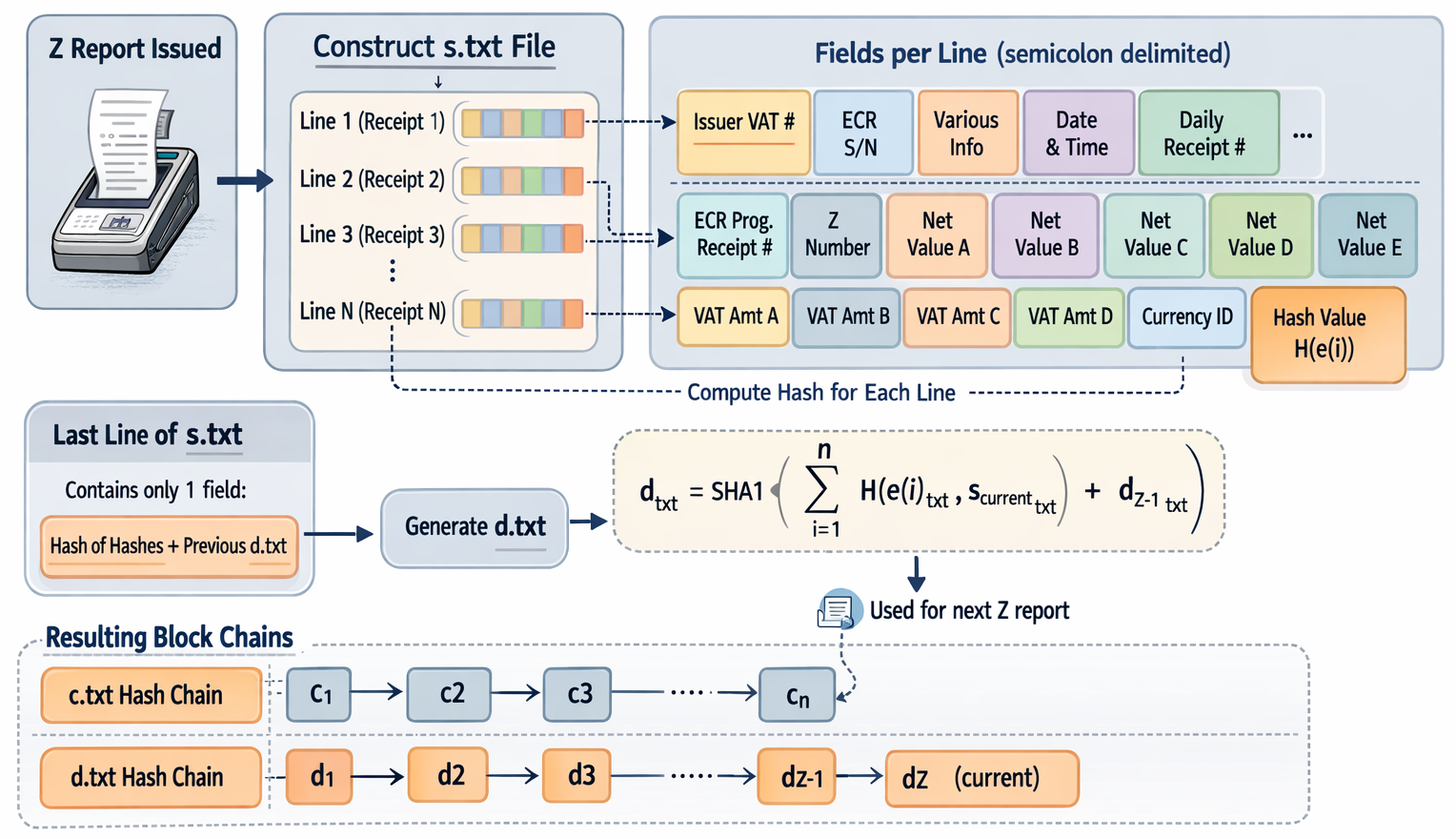}}
\caption{The blockchain logic for the double-hash architecture}
\label{fig}
\end{figure*}

\subsection{Fiscal Printers: Triple Blockchain Implementation}

The Greek Fiscal Printers (Fig.~\ref{fig3}) have the official name "Autonomous Fiscal Devices" “AFDs” (POL 1220/2012). In reality they are Fiscal ECRs without a keyboard. Instead of a keyboard, they accept sales orders from a computer according to a specific protocol.

Thus a fiscal printer does not print a random selected text, but responds only to specific commands according to the communication protocol e.g. item sale, discounts,  payment acceptance.
Apart from the lack of a keyboard, all other functions and features of the fiscal printers are the same as those of the Fiscal ECRs.  The implementation of the triple block chain is identical to that of Fiscal ECRs.

\subsection{Fiscal Signing Machines - EAFDSS: Double Blockchain Implementation}

As described in POL 1220/2012, the functionality of  Greek Fiscal Signing Machines “EAFDSS” is the following: 
The EAFDSS is connected to a computer (e.g. using the COM1 port). The printer where fiscal documents are printed, is connected to another port of the same computer (e.g. the LPT1 port). So the printer and the EAFDSS are not connected in tandem and the printer is not fully dedicated to the EAFDSS. As a side effect, it is very easy for any ERP running in the computer, to bypass the EAFDSS and print a fake fiscal document directly to the printer used by the EAFDSS.

\begin{figure}[htbp]
\scalebox{.25}{\includegraphics{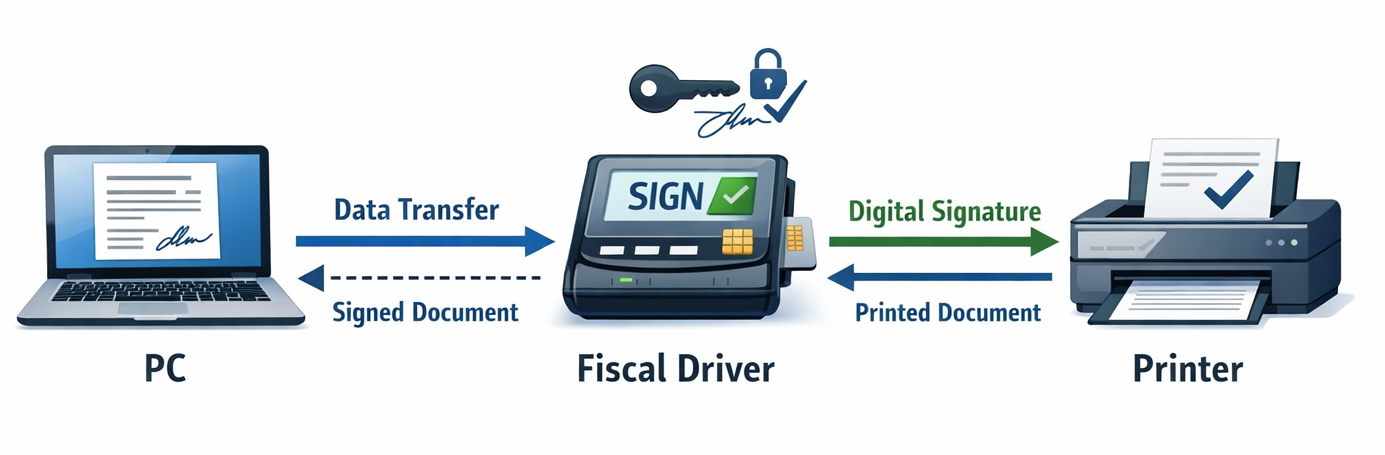}}
\caption{EAFDSS connectivity.}
\label{fig}
\end{figure}

The EAFDSS is always accompanied by a driver that is running as a daemon in the connected computer. When any document is going to be printed, the EAFDSS driver parses the document and diverts it to the EAFDSS. The EAFDSS functions as an SHA-1 algorithm \cite{b16} and computes a 40 Hex sequence as a hash value of the input document. Upon completion of these functions, three files are constructed with names ending in “a.txt”, “b.txt”, “e.txt” for the specific document. The first file “a.txt” contains the original document in plain text format. The second file “b.txt” contains the hash value that corresponds to the “a.txt”  file. The third file “e.txt” contains the  fiscal values of the following fields,  described in table C of POL 1220/2012 : issuer VAT number, receiver VAT number,  ECR s/n, various information, date and time, daily receipt/invoice number, EAFDSS receipt/invoice  progressive number, Z number, invoice type, invoice order, invoice number, net value A, net value B, net value C, net value D, net value E, VAT amount A, VAT amount B, VAT amount C, VAT amount D, grand total, currency id, hash value.
Note that the last field computed, is another hash value of 40 Hex, derived on the other fields, using the SHA-1 algorithm. Note also that the EAFDSS “e.txt” fields are a superset of the ECR/ADHME “e.txt” fields, so a unique central server can be used to store all receipts and invoices issued by either ECR/ADHME/EAFDSS.

Upon issuing a “Z” report, a new 40 hex hash value using the SHA-1 algorithm is computed, over the previous Z hash value and all b.txt hash values, resulting a file named “c.txt”.\\

\begin{equation}
EADFSS-c_{\text{txt}} =
\mathrm{SHA1}\!\left(
\sum_{i=1}^{n} H(b_{\text{txt}}^{(i)}, Z_{\text{txt}}^{\text{current}})
\;+\;
c_{\text{txt}}^{Z-1}
\right)
\label{eq:dtxt}
\end{equation}

This hash value is also stored in a fiscal memory entry.
In this a way, a block chain of hash values is created, which is stored to files “c.txt” and also into the fiscal memory entries. Note that early Greek EAFDSS, as described in POL 1135/2005, as they did not produce “e.txt” files, performed only this block chain implementation. 
New Greek EAFDSS, as described in POL 1220/2012, implement an additional block chain. The last line of the “s.txt” file consists only from one field, that is the hash value of the hash values of the previous lines plus of the total hash value of  the “s.txt” file of the previous “Z” report.  This hash value is referred as  “d.txt”.\\

\begin{equation}
EADFSS-d_{\text{txt}} =
\mathrm{SHA1}\!\left(
\sum_{i=1}^{n} H(e_{\text{txt}}^{(i)}, s_{\text{txt}}^{\text{current}})
\;+\;
d_{\text{txt}}^{Z-1}
\right)
\label{eq:dtxt}
\end{equation}

In this way, a double block chain of hash values is constructed, one with the “c.txt” hash values  and another with the “d.txt” hash values.

\subsection{Fiscal Signing Machines - FEMAS: Triple Blockchain Implementation}

Greek new Fiscal Electronic Mechanisms of Accumulation and Signing “FEMAS” are defined in A.1173/2022.  A FEMAS accommodates the characteristics of a Fiscal Electronic Cash Register and of a Fiscal Signing Machine “EAFDSS” in one device. A FEMAS operates within a tampered case sealed by a fiscal screw, which allows access only to special authorized technicians. A FEMAS has a working memory, an integrated SD card and a fiscal memory. The fiscal memory is typically an EPROM, or other type of memory controlled by a microprocessor, so that only writing new data is allowed, while deleting or modifying data is prohibited. The fiscal memory, for greater security, is located within a special location, and is surrounded by a special resin.
In the working memory of the FEMAS, the net value and VAT amounts of each issued receipt are added up. When a 'Z' report is issued, these amounts are transferred from the accumulators to a new entry in the fiscal memory so that the working memory can accommodate the new data.
In addition to the above memories, each FEMAS has a built-in SD card, where each receipt is recorded. The original text of the receipt is placed in a file with the extension a.txt which is recorded on the SD card.  When a receipt is issued, a hash value shall be calculated using the SHA-1 algorithm, which shall be recorded in a corresponding file with the extension b.txt on the SD card. The FEMAS also creates a file with the extension e.txt which is also recorded on the SD card. The e.txt file consists of specific fields of fiscal interest defined in Table C of POL 1220/2012.
Therefore, the basic mode of operation is similar to that of the Fiscal Signing Machine “EAFDSS” described earlier, creating for each receipt, files with the extension a.txt, b.txt, e. txt, with the difference that these files are stored on the built-in SD card of the FEMAS and not on the hard disk of the connected computer as in the case of the Fiscal Signing Machine “EAFDSS”. In this way, access to the FEMAS’s  files a.txt, b.txt, e.txt is impossible.
A FEMAS contains additional registers, accumulators and totalizers similar to those of Fiscal ECRs. During the procedure of issuing a receipt, the FEMAS separates the data of fiscal interest described in Table C of POL 1220/2012 and if they are amounts, adds them to the sums of  the corresponding amount accumulators.
When a Z report is issued, a c.txt file is created, which is a hash value derived using the SHA-1 algorithm, of the hash values of the current Z's  b.txt files  and of  the hash value  of the previous Z.
Z=c.txt=SHA1{ all b.txt of current Z + c.txt of previous Z}
This hash value is also saved  into the fiscal memory.
As mentioned earlier in the description of the “EAFDSS”, upon issuing a Z report, a file s.txt is also created, consisting of the lines of the e.txt of the current Z and a line d.txt which contains a hash value resulting from the hash values  of the e.txt lines of the current Z and the line d.txt of the previous Z.
d.txt=SHA1{ all hash values of e.txt of current s.txt  + d.txt of previous Z }.
In this way, a double block chain has been created, in exactly the same way as it is created in the Fiscal Signing Machines “EAFDSS”.
Finally, when the Z is issued, the contents of the accumulators of the net value amounts and of the VAT amounts are transferred to the corresponding entry in the  fiscal memory. In the same fiscal memory entry, the new progressive totals are calculated as follows:
Progressive total of current Z = Total value of current Z + progressive total of previous Z
In this way, another block chain is implemented, like the one implemented in Fiscal ECRs for monitoring progressive net amounts and progressive VAT amounts. A FEMAS therefore implements a triple block chain.

\section{Connection of EFTPOS and Greek Fiscal Electronics Mechanisms: Blockchain Implementation of EFTPOS Transactions}

Electronic Funds Transfer at Point of Sale “EFTPOS” is a convenient way for customers to pay using one of many payment methods – usually debit or credit cards, mobile phones, smart watches, or other wearables.

\begin{figure}[htbp]
\centerline{\includegraphics{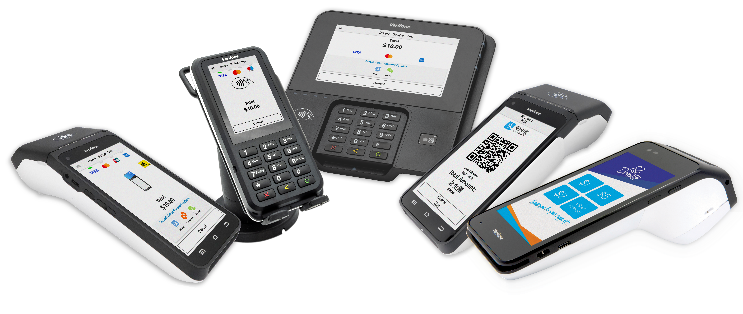}}
\caption{Various EFTPOS types.}
\label{fig}
\end{figure}

According to the EFTPOS-FEM communication protocol described in A.1098/2022 \cite{b18}  as amended by A.1088/2023 \cite{b19}, A.1027/2024 \cite{b20}, EFTPOS operating in Greece must be connected to Fiscal Electronic Mechanism "FEM".  The purpose of this interconnection is to prevent the use of EFTPOS for accepting payments, if the receipt has not been previously issued by FEM. For this reason, the EFTPOS keypad is locked and does not allow entering an amount. The amount of the transaction is transmitted to EFTPOS by the FEM only after the fiscal receipt has been issued.
In this way, all EFTPOS transactions e.g. payments by credit card, must be paired by corresponding fiscal receipts. Before the EFTPOS transaction, the following data are sent to the EFTPOS : session number, FEM id, amount. After each EFTPOS transaction, the following data are sent to the FEM:
Session number \# Acquirer id \# terminal id \# rrn   \# date and time \# amount\_final \# txn-ecr-status. These data are printed on the FEM printer, immediately after printing the fiscal ECR receipt.
Note that the amount final may be lower than the amount specified in the fiscal receipt, as various discounts may apply, e.g. due to loyalty. In any case, FEM sums up the card payment amounts in a special accumulator in the working memory and prints the total amount of card payments when issuing the Z report. It shall also print the progressive card payment amount when reporting Z is issued, which shall also be stored in a special record in the fiscal memory. This creates a blockchain for the progressive card payment totals in EFTPOS.
In addition, when issuing Z, FEM transmits the card payments received from EFTPOS in detail to the Central Fiscal Electronic Mechanisms database “eSEND”. In this way, the tax and customs auditors of the IAPR can check the card payments and determine whether they are matched with the issued retail fiscal receipts.

\section{The Central Information System eSEND"}

In the event that there is no internet connection when the Z report is issued, then in accordance with the provisions of POL 1166/2018 \cite{b21}, the FEM keeps the entire Z sales data in its internal SD card memory and will send them the next time a Z is issued. In this way, the FEM ensures that no sales data will be missing for any receipt.
The transmission data of each FEM is encrypted using the AES-256 symmetric algorithm. The AESkey is unique for each FEM and is sent to each FEM with a call to an eSEND web service, in accordance with the provisions of A.1088/2023. This AESKey is used by the FEM to encrypt the packages sent to "eSEND" and by “eSEND” to decrypt the packages from that specific FEM.
The FEM sales data are sent to the central information system "eSEND" of the IAPR. This central information system consists of 2 parts:

\subsection{The back-end: Receiving sales data from the FEMs}\label{AA}	
The back-end process of the central information system ``eSEND'' runs continuously in the Greek General Secretariat for Information Systems``GSIS'' and it is the one to which the FEMs are automatically linked when they issue the ``Z'' report.  According to POL1166/2018, A.1024/2020, A.1173/2022, the central information system ``eSEND'', after determining that it is a valid FEM, receives the encrypted data, which it decrypts using the AESkey related to the specific FEM. The central information system ``eSEND'' has stored the information up to which Z number each FEM has sent data, so it allows to accept data from the next Z number onwards. If for example ``eSEND'' has accepted data up to Z \#45, the next data it expects will be of Z \#46, then Z \#47 and so on. 

Furthermore, ``eSEND'' performs a check, if the d.txt line of each Z, follows the rule of the blockchain described in sections II,III,IV,V above. If the blockchain rule is not followed, the Z sales data are discarded.

In this way, it is impossible for any Z's data to be missing and ``eSEND'' achieves an integrity of received sales data.
The data obtained for each Z report is detailed by each fiscal receipt or invoice issued.  The data are entered in a table in the IAPR central database for further use. The data fields are defined in table C of POL 1220/2012 as amended by A.1173/2022 as follows:

\begin{table}[htbp]
\caption{Table C - transmitted Data Fields to eSEND}
\begin{center}
\begin{tabular}{|c|c|}
\hline
\textbf{Table C} & \textbf{\textit{Field Length}} \\
\hline
VAT  no. of issuer& (12)  \\
VAT  no. of recipient& (12)  \\
Serial Number of FEM& (12) \\
Additional information & (200)  \\
Date and time (YYYYMMDDHHmm) & (12)\\
Daily Serial number of sales document&(5)\\
Progressive Serial number of sales document&(5)\\
Z number&(4)\\
Code (type) of Sales Document&(4)\\
Series&(10)\\
sales Document number&(10)\\
Net Amount A&(N-18:2)\\
Net Amount B&(N-18:2)\\
Net Amount C&(N-18:2)\\
Net Amount D&(N-18:2)\\
Net Amount E&(N-18:2)\\Net Amount A&(N-18:2)\\
VAT A&(N-18:2)\\
VAT B&(N-18:2)\\
VAT C&(N-18:2)\\
VAT D&(N-18:2)\\
Grand Total of Sales Document&(N 18:2)\\
Currency code&(1)\\
NUMBER OF PAYMENTS BY CARD&(N:2)\\
TOTAL CARD PAYMENTS&(N 18:2)\\
CARD PAYMENT ID&(60)\\
MyData&(400)\\
Random&(10)\\
ASEDS&(40)\\

\hline
\multicolumn{2}{l}{$^{\mathrm{a}}$This table depicts the transmitted data fields schema.}
\end{tabular}
\label{tab1}
\end{center}
\end{table}

We would like to point out that regardless of the type of FEM, each sales document  ( \,invoice or retail receipt ) \, is accompanied by two serial numbers: the daily serial number for the specific Z and the progressive serial number regardless of Z. Therefore, the primary key of each sales document issued by a FEM is the combination of the FEM serial number and the progressive serial number.

\subsection{The Front-end: User Interface of the central information system “eSEND”}

\begin{figure}[htbp]
\centerline{\includegraphics{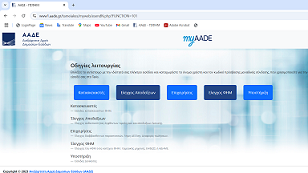}}
\caption{The central information system “eSEND” homepage in 2023.}
\label{fig6}
\end{figure}

The purpose of the Front-end of the central information system “eSEND”, is to display the sales data received from the FEMs. The Front-end of the central information system “eSEND” is a website, where every business and every citizen has access. This website is located at: www1.aade.gr/tameiakes/ (alias www1.gsis.gr/tameiakes/ ) .  
Since May 2023, the home webpage of the central information system changed and the UI is depisted in Fig.~\ref{fig6}.

In summary, we note that the following rules are followed for sending transaction data from FEMs to eSEND: \\
	-	Encryption of sales data with AES.\\
	-	Data transmission when  the Z report is issued,  provided that there is internet connection and eSEND is operating normally.\\
	-	If there is no internet connection or if there is any malfunction, then the FEM keeps the sales data in internal memory until operation is restored.\\
	-	If  a Z report is transmitted out of sequence, eSEND rejects it in order to receive the Z report in sequence.\\
	-	If the transmission data does not implement the blockchain, it is rejected by eSEND.\\
	-	 For each FEM, all sales document progressive numbers must be present one after the other.\\

\section{Functionality of Electronic Invoicing Provider Service}
\begin{figure}[htbp]
\scalebox{.2}{\includegraphics{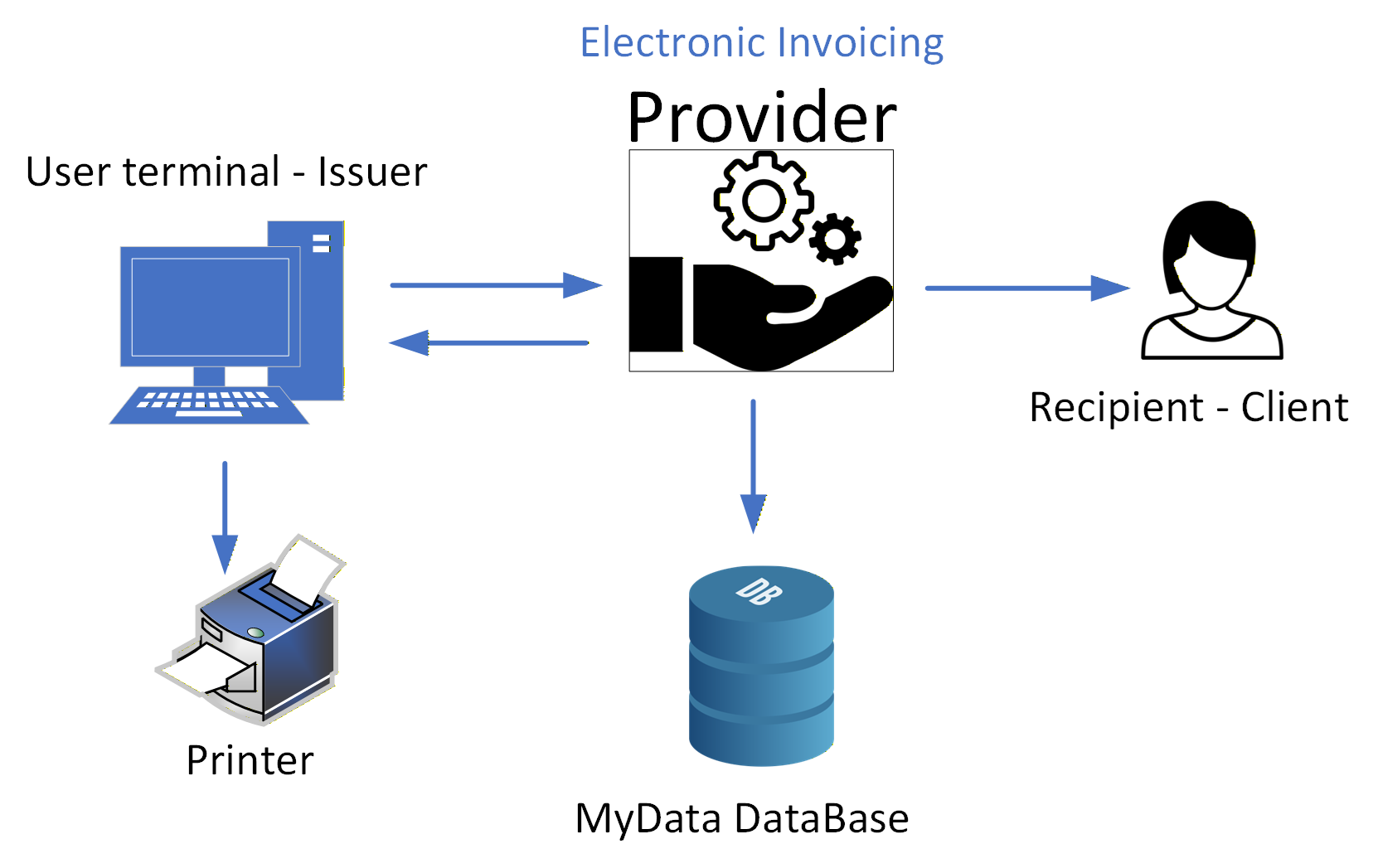}}
\caption{Functionality of an Electronic Invoicing Provider Service.}
\label{fig7}
\end{figure}

The obligations and the functionality of an Electronic Invoicing Provider Service ``EIPS'' are described in \cite{b22} and in the decision  A.1035/2020 \cite{b23}, as amended by A.1158/2023 \cite{b24}, A.1126/2024 \cite{b25} and by A.1112/2025 \cite{b26}. In the case where a receipt issuer uses the services of an Electronic Invoicing Provider ``EI'', the functionality is described in Fig.~\ref{fig7}.

When the receipt or invoice is issued, the issuer transmits the document via the EIPS application to the Electronic Invoicing Provider, which, after authenticating it using a hash algorithm, then transmits it to the central database "myData" of the Independent Authority of Public Revenue ``IAPR'' and also to the recipient or to the recipient's Electronic Invoicing Provider. It is clarified that the transmission to the recipient is accomplished provided that the recipient has the infrastructure to receive electronic documents. Finally, in the case of a retail transaction, the issuer prints the retail receipt on the local printer.

\section{Rules for Data Transfer from EIPS to MyData}

Provided there is an internet connection and the issuer is connected to the e-envoicing provider, the sales document is transferred from the issuer's computer to the e-envoicing provider in real time and from there to the central database 
“myDATA”. All data are transmitted using https protocol. 
However, in cases where there is no connection between the issuer and the electronic invoicing provider, e.g. due to internet malfunction, then according to A.1035/2020, as amended by A.1112/2025, the issuer has the right to issue receipts or invoices offline for 24 hours and then transmit them to the e-invoicing provider. However, the absence of a FEM with HW safeguards, but instead the use of a simple computer that allows data to be deleted or sales documents to be modified before transmission, allows the trader-issuer to delete or modify the sales documents before transmitting them to the electronic invoicing provider.
The issuer has the right to restart the numbering of sales documents from the beginning at the start of each fiscal period. This means that the document's serial number is not the primary key.
Both in the process of issuing documents via the EIPS and in the process of transmitting documents to the central database of the IAPR "myData", the implementation or the maintenance of a blockchain [27] is not required. Therefore, it is allowed to issue the invoice with serial number 1000, without having previously issued the invoice with serial number 999. Thus, there is no integrity check.
In response to the transmission of a sales document, myDATA replies with a unique MARK number. However, while these numbers are unique for the central myDATA database, they are not consecutive for two consecutive sales documents from the same business. Therefore, when looking at a company's account in myDATA, the tax auditor cannot conclude from the MARKs whether all sales documents have been sent consecutively to myDATA or whether some are missing.

\section{Conclusion}

Using tamper-proof hardware, fiscal memories, unique progressive sales document numbers, AES encryption, and blockchain, FEMs transmit all the sales documents they issue to the central information system ``eSEND''. After decrypting each Z package containing detailed sales data, ``eSEND'' checks whether the Z number is the next one after the Z number that has already been transmitted, and whether the hash value of the Z package verifies the blockchain rule. In this way, “eSEND” ensures sales data integrity.
On the contrary, in the case of EIPS that transmit sales data to the central information system ``myDATA'', the absence of tamper-proof hardware, the absence of fiscal memories, the absence of progressive sales document numbers, the absence of consecutive MARK numbers for consecutive sales documents of the same business, the absence of a blockchain, and the absence of a requirement to transmit a sales document only if the immediately preceding one has been transmitted to ``myDATA'', leads to a lack of sales data integrity in ``myDATA." It also leads to doubt on the part of tax auditors as to whether the audited company has transmitted to ``myDATA'' all of its sales documents or only a part of them.

\vspace{12pt}
\color{red}


\begin{thebibliography}{00}

\bibitem{b1}
M. Iansiti and K. R. Lakhani, “The Truth About Blockchain,” 
\textit{Harvard Business Review}, vol. 95, no. 1, pp. 118--127, 2017.

\bibitem{b2}
S. Nakamoto, “Bitcoin: A Peer-to-Peer Electronic Cash System,” 2008. [Online]. Available: https://bitcoin.org/bitcoin.pdf

\bibitem{b3}
M. Swan, \textit{Blockchain: Blueprint for a New Economy}. Sebastopol, CA, USA: O’Reilly Media, 2015.

\bibitem{b4}
J. S. Sogaard, “A Blockchain–Enabled Platform for VAT Settlement,” 
\textit{International Journal of Accounting Information Systems}, p. 100502.

\bibitem{b5}
K. Yent, “Applying Blockchain Technology to Cross-Border Tax Reporting,” University College London Centre for Blockchain Technologies, Discussion Paper no. 4, 2020.

\bibitem{b6}
G. Bossa and E. de Paiva Gomes, “Blockchain: Technology as a Tool for Tax Information Exchange or an Instrument Threatening the Taxpayer’s Privacy?,” SSRN Electronic Journal, 2019. [Online]. Available: https://ssrn.com/abstract=3540277

\bibitem{b7}
E. Mik and N. Noked, 
“Blockchain and Tax Administration: A Critical Assessment.”

\bibitem{b8}
C. J. Delmotte, 
“Toward a Blockchain-Driven Tax System.” [Online]. Available: https://ssrn.com/abstract=4187919

\bibitem{b9}
R. T. Ainsworth and M. Alwohaibi, 
“Blockchain, Bitcoin and VAT in the GCC: The Missing Trader Example.” [Online]. Available: http://www.bu.edu/law/faculty-scholarship/working-paper-series/

\bibitem{b10}
“Council Directive 2006/112/EC,” 
\textit{Official Journal of the European Union}, L347, Dec. 11, 2006. [Online]. Available: https://eur-lex.europa.eu/legal-content/EN/TXT/PDF/?uri=CELEX:32006L0112

\bibitem{b11}
“Law 4308/2014: Greek Accounting Standards, related regulations, and other provisions,” 
\textit{Official Journal of the Greek Government}, no. 251A, Nov. 24, 2014, pp. 7651--7746. [Online]. Available: https://www1.aade.gr/tameiakes/myweb/Law-4308-2014-ENGLISH.pdf

\bibitem{b12}
P. Zafeiropoulos, 
“Completion of Technical Specifications of Fiscal Electronic Machines and Systems. Procedures for Their Use and Operation,” 
POL 1135/2005, \textit{Greek Government Gazette}, no. 1592B, Nov. 17, 2005.

\bibitem{b13}
P. Mavridis, 
“POL 1220/2012: Codification and Completion of Technical Specifications of Fiscal Electronic Mechanisms,” 
\textit{Greek Government Gazette}, no. 351B, Dec. 31, 2012. [Online]. Available: https://www1.aade.gr/tameiakes/myweb/POL1220-2012-ENGLISH.pdf

\bibitem{b14}
P. Mavridis, 
“A.1024/2020: Supplementary Technical Specifications of Communication and Encryption Protocol for the Transmission of Data to the FEM Information System,” 
\textit{Greek Government Gazette}, no. 317B, Feb. 6, 2020. [Online]. Available: https://www1.aade.gr/tameiakes/myweb/A1024-2020-ENGLISH.pdf

\bibitem{b15}
N. Stefos, J. Roussos and P. Mavridis, 
“A.1173/2022: Amendment of the Technical Specifications of Fiscal Electronic Mechanisms,” 
\textit{Greek Government Gazette}, no. 6593B, Dec. 30, 2022. [Online]. Available: https://www1.aade.gr/tameiakes/myweb/A1173-2022-ENGLISH.pdf

\bibitem{b16}
D. Hantzigianni and P. Mavridis, 
“POL 1068/2015: Procedures for the Approval and Revocation of Fiscal Electronic Mechanisms,” 
\textit{Greek Government Gazette}, no. 497B, Apr. 1, 2015. [Online]. Available: https://www1.aade.gr/tameiakes/myweb/POL1068-2015-ENGLISH.pdf

\bibitem{b17}
“SHA-1.” [Online]. Available: https://en.wikipedia.org/wiki/SHA-1

\bibitem{b18}
“A.1098/2022: Determination of Technical Specifications for the Interconnection between Fiscal Electronic Mechanisms and EFT/POS Terminals,” 
\textit{Greek Government Gazette}, no. 3940B, Jul. 25, 2022. [Online]. Available: https://www1.aade.gr/tameiakes/myweb/A1098-2022-ENGLISH.pdf

\bibitem{b19}
P. Mavridis, 
“A.1088/2023: Amendment of Decision A.1098/2022 on the Interconnection between Fiscal Electronic Mechanisms and EFT/POS Terminals,” 
\textit{Greek Government Gazette}, no. 3875B, Jun. 15, 2023. [Online]. Available: https://www1.aade.gr/tameiakes/myweb/A1088-2023-ENGLISH.pdf

\bibitem{b20}
Independent Authority for Public Revenue (IAPR),
``Amendment of the Decision of the Governor of IAPR A.1098/13-7-2022: Determination of Technical Specifications (Communication Protocol) and Other Functions for the Interface Between Fiscal Electronic Mechanisms and EFT/POS Terminals. Implementation of the Principle of "Collection by Card – Mandatory Issuance of a Receipt by a FEM",''
Decision A.1027/2024, Government Gazette (FEK) B' 1134, Feb. 16, 2024.

\bibitem{b21}
Independent Authority for Public Revenue (IAPR),
``Technical Specifications of Communication and Encryption Protocol for the Transmission of Data to the FEM Information System,''
Decision POL.1166/2018, Government Gazette (FEK) B' 3603, Aug. 24, 2018. Available: https://www1.aade.gr/tameiakes/myweb/POL1166-2018-ENGLISH.pdf

\bibitem{b22}
P. Mavridis, A. T. Baklezos, and C. D. Nikolopoulos,
“e-Invoicing Provider Services in Greece: New Perspectives and Weaknesses,”
in \textit{Proc. 5th Int. Conf. Electronic Engineering, Information Technology and Education (EEITE)},
Athens, Greece, 2024, pp. 1--4, doi:10.1109/EEITE61750.2024.10654410.

\bibitem{b23}
“A.1035/2020: Obligations of Electronic Invoicing Providers and Control Procedures for the Provision of Electronic Invoicing Provider Services,” 
\textit{Greek Government Gazette}, no. 551B, Feb. 20, 2020. [Online]. Available: https://www.vatupdate.com/wp-content/uploads/2020/02/2020-02-21-Greece-Einvoicing-English.pdf

\bibitem{b24}
“A.1158/2023: Amendment of Decision A.1035/2020 on Electronic Invoicing Provider Services,” 
\textit{Greek Government Gazette}, no. 5996B, Oct. 13, 2023. [Online]. Available: https://www1.aade.gr/tameiakes/myweb/A1158-2023-ENGLISH.pdf

\bibitem{b25}
“A.1126/2024: Amendment of A.1035/2020 on Electronic Data Issuance Services,” 
\textit{Greek Government Gazette}, no. 4557B, Aug. 6, 2024. [Online]. Available: https://www1.aade.gr/tameiakes/myweb/A1126-2024-ENGLISH.pdf

\bibitem{b26}
“A.1112/2025: Amendment of Electronic Invoicing Provider Services,” 
\textit{Greek Government Gazette}, no. 4206B, Aug. 1, 2025. [Online]. Available: https://www1.aade.gr/tameiakes/myweb/A1112-2025-ENGLISH.pdf

\end{thebibliography}
\end{document}